\documentclass[aps,twocolumn]{revtex4}
\usepackage{graphicx}
\usepackage{dcolumn}
\usepackage{amssymb}

\begin{document}

\title{Detecting local synchronization in coupled chaotic systems}
\author{L. Pastur, S. Boccaletti and P.L. Ramazza}
\affiliation{Istituto Nazionale di Ottica Applicata, Largo E.
Fermi 6, 50125 Florence, Italy}
\date{\today}

\begin{abstract}
We introduce a technique to detect and quantify local functional
dependencies between coupled chaotic systems. The method estimates
the fraction of locally syncronized configurations, in a pair of signals
with an arbitrary state of global syncronization.
Application to a pair of interacting R\"ossler oscillators shows
that our method is capable to quantify the number of dynamical
configurations where a local prediction task is possible, also in
absence of global synchronization features.

\vglue 0.4 truecm


\end{abstract}

\maketitle

In the past years much attention has been devoted to characterize
coupled chaotic systems exhibiting synchronization regimes
\cite{reviews}. In this framework, different synchronization
features have been studied, such as, e.g., identical and
generalized synchronization \cite{complete,generalized}, phase
synchronization \cite{phase}, lag and intermittent lag
synchronization \cite{intermittentlag}. Furthermore,
synchronization effects have been explored in natural phenomena
\cite{nature}, and controlled laboratory experiments
\cite{experiment}.

In this context, various attempts to provide unifying definitions
for encompassing the different synchronization phenomena have been
pursuited \cite{previous}. Recently, a formal approach to the
problem has been put forward \cite{definition}, in which the
unifying property of synchronization is established in the
emergence of local functional dependencies between neighborhoods
of particular phase space configurations in the projected spaces
of the two coupled subsystems. The approach assumes a system {\bf
Z} $\in {\Bbb R}^{m_1+m_2}$ divisible into two coupled subsystems,
{\bf X} $\in {\Bbb R}^{m_1}$ and {\bf Y} $\in {\Bbb R}^{m_2}$. In
this framework, synchronization is equivalent to predictability of
one subsystem's values from another, \textit{i.e.} that an event
$\tilde y$ in {\bf Y} always occurs when a particular event
$\tilde x$ in {\bf X} occurs. However, when searching for evidence
of synchronization in data, one seldom has data that fall right on
a given $\tilde x$ or on a given $\tilde y$. Rather, the closer
$x(t)$ is to $\tilde x$ the closer $y(t)$ is to $\tilde y$. The
latter statement is captured rigorously by a local {\em
continuous} function; namely, the trajectories of $x(t)$ close to
$\tilde x$ are mapped near to $\tilde y$ by a local function that
is continuous at the point $(\tilde x, \tilde y)$, and that, near
($\tilde x, \tilde y$) describes well the predictability of
subsystem {\bf Y} dynamics from subsystem {\bf X} dynamics. Ref.
\cite{definition} gives a general, formal mathematical ground to
the above statements, and establishes the sufficient conditions
for a system to display global synchronization features,
\textit{i.e.} to admit local functional dependencies regardless on
the particular choice of the ($\tilde x, \tilde y$) phase space
configuration.

For a generic pair of coupled chaotic systems, however, it is to
be expected that synchronization occurs only at some locations (if
any) of the phase space, and not globally. In this case, a
continuous functional dependence of $y(t)$ on $x(t)$ will exist
only locally around a set of synchronization points $\{ \tilde
x_s, \tilde y_s \}$.

Implementation of a search for local functional dependencies
requires two separate steps: a preliminary one in which the two
interacting subsystems {\bf X} and {\bf Y} are properly identified
within the original dynamical systems {\bf Z}, and their
dimensionalities measured, and a second one in which the local
synchronization points ($\tilde x, \tilde y$) are detected. The
first problem was solved recently in Ref. \cite{disentangle} by
means of a modification of the {\it false nearest neighbors}
algorithm \cite{abarbanel}, allowing for a separate measurement of
the dimensionalities of weakly coupled systems in the case of
emergent synchronization motions.

In this paper, we will address the second step of the search by
introducing the {\it synchronization points percentage} ({\it
SPP}) indicator, and show how one can gather information on local
synchronization properties emerging in coupled chaotic systems.

We start by assuming to have $N$ data points in {\bf Z} $\in {\Bbb
R}^{m_1+m_2}$. By means of a proper subspace reconstruction
\cite{disentangle}, we end up with $N$ data points in {\bf X} $\in
{\Bbb R}^{m_1}$ and $N$ corresponding images in {\bf Y} $\in {\Bbb
R}^{m_2}$. We then pick a specific point $\tilde x \in$ {\bf X}
and consider its image $\tilde y \in$ {\bf Y}.

The first task consists in identifying proper domains and
co-domains for a statistical analysis of the existence of
functional dependency. For this purpose, we choose a pair of
positive real numbers $(\varepsilon_k, \delta)$ (the index $k$
being an integer), and consider the volume $U_{\varepsilon_k}
\subset$ {\bf X} ($V_\delta \subset$ {\bf Y}) containing all
points whose $m_1$-distance ($m_2$-distance) from $\tilde x$
($\tilde y$) is smaller than $\varepsilon_k$ ($\delta$).
Furthermore, we look at all points in {\bf X} falling within
$U_{\varepsilon_k}$, and verify the imaging condition, that is we
ask ourselves whether or not all images of the points  in
$U_{\varepsilon_k}$ fall within $V_\delta$. If the answer is no,
we choose $\varepsilon_{k+1} < \varepsilon_k$, and repeat the
above procedure. If for all $k$ the imaging condition is not
satisfied, the task ends with the conclusion that no local
functional dependency exists in the vicinity of the chosen
configuration $(\tilde x,\tilde y)$. If, instead, for a given
$\tilde k$ the imaging condition is verified, the task ends with
the identification of a valid pair $(\varepsilon_{\tilde
k},\delta)$, over which one has to test for the existence of a
continuous functional relationship.

\begin{figure}[!h]
\includegraphics[width=80mm]{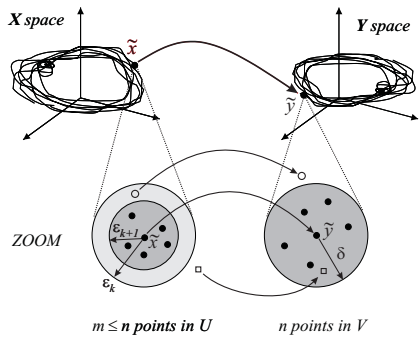}
\caption{Schematic representation of the statistical continuity
analysis. Upper part shows the reconstructed trajectories in the
two subspaces {\bf X} and {\bf Y}, and the location of the points
$\tilde x$ and $\tilde y$. The lower part zooms on the $U,V$
neighborhoods. For $\varepsilon=\varepsilon_{k+1}$, $V$ contains
all images of the $m$ points in $U$ (solid circles), plus images
of other points (empty squares) from outside $U$. For
$\varepsilon=\varepsilon_{k}$, some points in $U$ (empty circles)
have images outside $V$.}\label{fig1}
\end{figure}

Fig.\ref{fig1} helps in understanding the schematic representation
of the procedure. In the following we will denote with $U \subset$
{\bf X} ($V \subset$ {\bf Y}) the neighborhood
$U_{\varepsilon_{\tilde k}}$ ($V_\delta$) surrounding $\tilde x$
($\tilde y$), and assume that $m < N$ points fall within $U$. By
construction, the number of points falling within $V$ will be $n
\geq m$, reflecting the fact that $V$ might host also images of
points not belonging to $U$.

The probability of a  single point falling within $V$ is $P(V)
\equiv n/N$, and the probability that $m$ points fall within $V$
by pure chance is $P_m(V) = P(V)^m = \left( \frac{n}{N}
\right)^m$. This latter quantity, for reasonable choices of $n,m$
[reasonable pairs $(\varepsilon_{\tilde k},\delta)$], is a very
small number. However, one has to fix a {\it confidence level} of
comparison, for assessing existence of a local continuous function
between the two neighborhoods. This problem was addressed in
\cite{PecCarHea95}, where the {\it continuity statistics} method
was proposed. This consists in calculating the quantity $b_P$,
defined as

\begin{eqnarray}
\label{uno} b_P = \max_{q=1,...,m} B(q,m;P),
\end{eqnarray}

where $B(q,m;P)$ is the binomial distribution, giving the
probability that $q \leq m$ events out of $m$ attempts are
realized for a process of elementary probability $P$.

As said above, the presence of a single data within $V$ has
probability $P(V)$. The quantity $b_P$ (for $P=P(V)$) represents
then the maximum over $q$ of the probability that, given $m$
points, $q$ out of them fall into $V$. Hence, a level of
confidence for the existence of a continuous function can be
estimated in terms of the ratio

\begin{eqnarray}
\label{due} \Theta = \frac{P_m(V)}{b_P}.
\end{eqnarray}

If $\Theta \approx 1$ we have no trustable information about the
existence of such a functional relationship, insofar as the chance
probability of having our $m$ points in $V$ is of the same order
of the maximum probability of having events in $V$ out of $m$
attempts. On the contrary, if $\Theta \ll 1$, the chance
probability of having our $m$ points in $V$ is negligible compared
to $b_P$. Thus one concludes that the two sets $U$ and $V$ are the
domain and co-domain respectively of a local continuous function
mapping states in {\bf X} close to $\tilde x$ to states in {\bf Y}
close to $\tilde y$. This answers the practical question of
predicting states in {\bf Y} with error $\delta$ from measurements
of states in {\bf X} with error $\varepsilon_{\tilde k}$.

We have made use of the original formulation of the continuity
statistics \cite{PecCarHea95}, that explicitly considers $P=P(V)$
in Eq.(\ref{due}). More recently, the same Authors of
\cite{PecCarHea95} have proposed an alternative way for measuring
the confidence level, by choosing $P=1/2$ in the denominator of
Eq. (2), corresponding to an hypothesis of equal probability for
an attempt to fall within or outside the selected box
\cite{ECCpaper}.

Our technique for characterizing synchronization consists then of
the three following points: {\it i)} check the imaging of
neighborhoods of a given configuration $\tilde x$ into
neighborhoods of $\tilde y$; {\it ii)} assess the degree of
confidence that such an imaging process comes from the existence
of a local continuous function; {\it iii)} repeat points i) and
ii) for all $N$ pairs of configurations ($\tilde x, \tilde y$)
available in the data set. This procedure allows a classification
of the different dynamical states into locally synchronized and
non synchronized ones. As a result one can introduce the {\it
synchronization points percentage} ({\it SPP}) indicator, as the
ratio between the total number $\tilde n$ of locally synchronized
configurations and the total number of available points $N$.

The proposed method can be applied to any kind of multivariate
data set, for the detection of hidden local synchronization
properties, that cannot be detected by global indicators, such as
correlation functions, Lyapunov exponents, Lyapunov functionals,
or any other kind of time (or ensemble) average indicators that
unavoidably result in mixing locally synchronized and
unsynchronized configurations. As a result, the {\it SPP}
indicator furnishes relevant information in all those cases in
which synchronization states emerge locally in phase space, to
detect predictability properties that are limited to some subset
of the dynamics.

In order to illustrate the robustness of the method, in the
following we will refer to a test case, represented by a pair of
non identical bidirectionally coupled chaotic R\"ossler
oscillators. Here $m_1=m_2=3$, and the subspaces {\bf X} and {\bf
Y} contain state vectors {\bf x}$\equiv (x_1,y_1,z_1)$ and {\bf
y}$\equiv (x_2,y_2,z_2)$ whose evolution is ruled by

\begin{eqnarray}
\label{Rossler}
\dot{x}_{1,2}&=&-\omega
_{1,2}y_{1,2}-z_{1,2}+\epsilon ( x_{2,1}-x_{1,2}),\nonumber\\
\dot{y}_{1,2}&=&\omega _{1,2}x_{1,2}+0.165z_{1,2},\\
\dot{z}_{1,2}&=&0.2+z_{1,2}( x_{1,2}-10).  \nonumber
\end{eqnarray}

In Eqs.(\ref{Rossler}), $\omega _{1,2}=\omega _0 \pm \Delta$
represent the natural frequencies of the two chaotic oscillators,
$\omega _0=0.97$, $\Delta =0.02$ is the frequency mismatch and
$\epsilon>0 $ rules the coupling strength. As $\epsilon$
increases, the emergence of different synchronization features in
Eqs.(\ref{Rossler}) has been described and characterized in the
literature \cite{phase,intermittentlag}. Precisely, for $\epsilon
< 0.036$ no global synchronization (NS) is established, in terms
of the global indicators proposed up to now. For $0.036 \leq
\epsilon \preceq 0.11$ a phase synchronized (PS) regime emerges
characterized by the boundedness in time of the phase difference
$\Delta \phi \equiv \mid \phi_1 - \phi_2 \mid$ [$\phi_{1,2} \equiv
\arctan \left( \frac{y_{1,2}}{x_{1,2}} \right)$ being the phases
of the two oscillators], whereas the two chaotic amplitudes remain
almost uncorrelated \cite{phase}. At larger coupling strengths
($\epsilon \geq 0.145$), lag synchronization (LS) is established,
corresponding to a collective motion wherein $\mid \text{{\bf
x}}(t) - \text{{\bf y}}(t-\tau) \mid$ is bounded over the whole
dynamical evolution ($\tau>0$ represents here a lag time)
\cite{intermittentlag}. In this regime, increasing $\epsilon$
results in gradually decreasing $\tau$, eventually ending with a
regime indistinguishable from complete synchronization (CS).

Most of the transition points between these regimes were also
identified in Ref. \cite{intermittentlag}, by inspection of the
Lyapunov spectrum of Eqs.(\ref{Rossler}) as a function of the
coupling strength. Precisely, the NS to PS (PS to LS) transition
occurs for that value of $\epsilon$ for which a previously zero
(positive) Lyapunov exponent becomes negative. On the other hand,
the LS to CS transition is a smooth transition that can be tracked
by use of the time averaged similarity function
\cite{intermittentlag}. In the following we apply our method with
a threshold value of $\Theta =0.1$ for the discrimination of
whether or not the coupled systems display local functional
relationships.
\begin{figure}[!h]
\includegraphics[width=75mm]{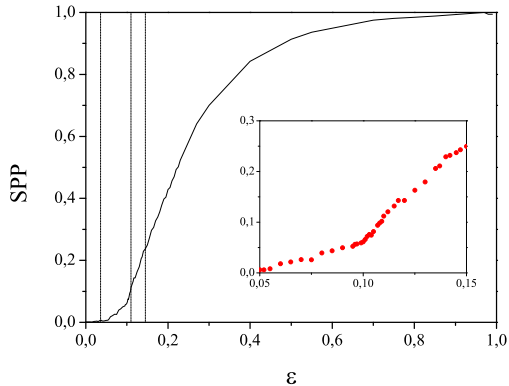} \caption{ {\it SPP} indicator (see
text for definition) {\it vs.} coupling strength $\epsilon$. The
vertical dashed lines indicate the transition points between the
different synchronization regimes. The inset shows a zoom limited
to the range $0.05 < \epsilon <0.15$, where the PS to ILS
transition point is located at $\epsilon_{c} \simeq 0.10$. Notice
the two different slopes in the linear growth of {\it SPP} for
$\epsilon<\epsilon_{c}$ and $\epsilon>\epsilon_{c}$.}\label{fig2}
\end{figure}

An intermediate synchronization regime between PS and LS exists in
the range $0.11 \preceq \epsilon < 0.145$, called intermittent lag
synchronization (ILS), where the system (\ref{Rossler}) displays
long epochs of LS evolution, interrupted by persistent bursts of
desynchronized motion. This has been observed numerically, and put
in relation with the system's trajectory passing through
configurations where one globally negative Lyapunov exponent has a
local positive value. Since ILS is an intimately local phenomenon,
its transition point has not been captured by those techniques
that measure time or ensemble averaged quantities. As a result, up
to now, studies on ILS have been limited to numerical
investigations \cite{intermittentlag}, or based upon the role in
the synchronization process played by the different unstable
periodic orbits visited by the dynamics \cite{pazozaks}. We will
show that our {\it SPP} indicator is able to discriminate between
ILS and PS regimes, as well as to directly identify the PS to ILS
transition point.

We have performed long time simulations of Eqs.(\ref{Rossler}) at
several coupling strength values, and collected data points from
the two scalar outputs $x_1$ and $x_2$. For each $\epsilon$, data
points are collected over a time corresponding to $1.7\cdot 10^5$
R\"ossler cycles, with a sampling frequency of 10 points per
cycle. Simulations were performed with a standard fourth order
Runge-Kutta method, and with random initial conditions.
Furthermore, the standard embedding technique \cite{takens} was
used to reconstruct the three dimensional vector states {\bf x}
and {\bf y} from time-delayed coordinates of the scalar variables
$x_1$ and $x_2$, and calculation of the {\it SPP} indicator was
performed on the reconstructed spaces.

\begin{figure}[!h]
\includegraphics[width=75mm]{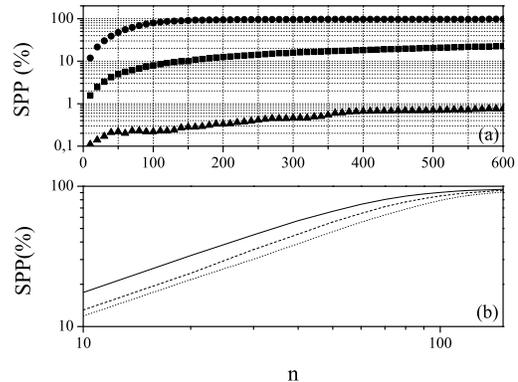}
\caption{a) {\it SPP} indicator {\it vs.} number $n$ of points
falling within $V$ for $\epsilon=0.01$ (triangles, NS),
$\epsilon=0.08$ (squares,PS) and $\epsilon=0.155$ (circles, LS).
b) {\it SPP} {\it vs.} $n$ within the LS regime for
$\epsilon=0.155$ (dotted line); $\epsilon=0.17$ (dashed line) and
$\epsilon=0.19$ (continuous line). For all cases, before
saturation ($n \leq 50$), {\it SPP} depends on $n$ with a scaling
law {\it SPP} $\sim n^\beta$ with $\beta \sim 0.85$. For $n > 50$
the three curves saturate to 100 \% of synchronization
points.}\label{fig3}
\end{figure}

Fig.\ref{fig2} reports the behavior of the {\it SPP} indicator
\textit{vs.} the coupling strength $\epsilon$, calculated by
fixing $\delta$ so as $n=150$ points are falling within $V$.
Fixing $n$ results in general in an error $\delta $ that is not
constant over the attractor. On the other side, if the measure is
strongly non homogeneous, fixing $\delta $ could generate
situations in which $n$ is so small that the statistics becomes
meaningless. These concerns do not apply however in the case of
the R\"ossler system for the parameters used in Eqs. (3), since
the density of points is roughly homogeneous over the attractor
and both choices lead to equivalent results. As one expects, {\it
SPP} increases monotonically as the coupling strength increases,
saturating to 1 when approaching the CS regime.

Interesting novel information can be extracted by inspection of
{\it SPP} within those synchronization regimes, such as PS and ILS
that do not correspond to global synchronization features. In
particular, it is found that {\it SPP} is linearly increasing with
$\epsilon$ in both regimes, but with two different slopes (see the
inset of Fig.\ref{fig2}). The linear increase of the indicator
already within PS is a relevant result. Indeed, if and to which
extent PS implies weak correlations in the chaotic amplitudes was
yet unknown, and constituted an issue generating controversy. The
present result shows that PS does imply an increasing percentage
of local functional relationship, thus quantifying directly the
degree of amplitude synchronization within such a regime.
Furthermore, the crossover point between the slopes of the two
linear growths allows one to identify the PS to ILS transition
point at $\epsilon  \simeq 0.10$, that none of the various
indicators used in previous works was capable to reveal.

Finally, other novel information can be extracted from the scaling
behavior of {\it SPP} with $n$, that is with enlarging the radius
$\delta$ of the image box in the {\bf Y} subspace.
Fig.\ref{fig3}a) shows {\it SPP} \textit{vs.} $n$ for the NS, PS
and LS regimes. In all cases, the {\it SPP} indicator increases
monotonically. For LS (circles) it fastly saturates to 1 (the same
value as CS). This is reflecting the fact that LS differs from CS
only due to the presence of a lag time $\tau$. Enlarging too much
the neighborhood size results in $V$ to fully overlap with all
images of points in $U$ shifted by a phase factor $\omega \tau$,
where $\omega$ is the mean frequency of the oscillator, thus
making indistinguishable LS from CS.

More insights on this property can be extracted from
Fig.\ref{fig3}b), where {\it SPP} is reported \textit{vs.} $n$
within the LS regime for different values of $\epsilon$,
corresponding to different values of the lag time $\tau$. Here one
sees that, before saturation, {\it SPP} depends on $n$ with a
scaling law {\it SPP} $\sim n^\beta$ with a unique exponent $\beta
\sim 0.85$ for the three $\epsilon$ values. However, the three
curves saturate to 1 at three different values of $n$, reflecting
the behavior of $\tau$ within LS, that monotonically decreases as
$\epsilon$ increases.

Coming again to Fig.\ref{fig3}a), one realizes that for both NS
(triangles) and PS (squares), the {\it SPP} indicator is always
bounded away from 1. This indicates that in these regimes a global
predictability of one subsystem's states from measurement in the
other subspace is never possible for any choice of resolution.
However, given a resolution $\delta$ in the image subspace (a
maximum error allowed in the prediction), our indicator quantifies
the number of states that can be locally predicted at that
resolution, thus revealing that local hidden synchronization
features can be extracted for prediction purposes, also in those
cases in which global dependencies are not established. This
feature might be relevant for detecting configurations where a
local prediction can be assessed, in many situations where a
global prediction procedure fails.

In real data, the effect of noise is to reduce the resolution in
the phase space, so that the statistics relative to boxes
containing a small number of points is not reliable anymore. A
threshold in $n$ should therefore be introduced, typically
corresponding to $\delta$'s larger than the noise-induced
uncertainty.

Authors are indebted with L. Moniz and L.M. Pecora for many
fruitful discussions. Work partially supported by EU Contract
HPRN-CT-2000-00158, and MIUR Project FIRB n.  RBNE01CW3M\_001.
L.P. acknowledges support from contract MCFI-2000-01822.

\end{document}